\documentclass[10pt]{article}
\usepackage[OE]{express}
\usepackage{textcomp}
\usepackage{siunitx}
\usepackage{cite}

\begin{document}

\title{Experimental demonstration of high sensitivity refractive index sensing based on magnetic plasmons in a simple metallic deep nanogroove array}

\author{D. M. Li \authormark{1 $\dagger$}, X. Y. Kuang\authormark{1 $\dagger$}, H. Zhang\authormark{2,3}, Y. Z. Liang \authormark{4}, T. Xu\authormark{4}, L. Y. Qing\authormark{2},Y .H. Zhu \authormark{2}, S. Zhang\authormark{2,4}, W. X. Wang\authormark{5}, and W. Wang\authormark{2 *}}

\address{\authormark{1}Institute of Atomic and Molecular Physics, Sichuan University, Chengdu 610065, China\\
	
	\authormark{2}College of Physical Science and Technology, Sichuan University, Chengdu 610064, China\\
	
	\address{\authormark{3}Key Laboratory of High Energy Density Physics and Technology of Ministry of Education, Sichuan University, Chengdu 610065, China
		
	\authormark{4}National Laboratory of Solid State Microstructures, College of Engineering and Applied Sciences and Collaborative Innovation Center of Advanced Microstructures, Nanjing University, 210093, Nanjing, China}

	\authormark{5}Institut f\"ur Physik $\&$ IMN MacroNano (ZIK), Technische Universität Ilmenau, 98693 Ilmenau, Germany\\
}

\email{\authormark{*}w.wang@scu.edu.cn} %% email address is required
\email{\authormark{$\dagger$}These two authors contribute equally to this study} %% email address is required

%%%%%%%%%%%%%%%%%%%%

%\author{Guilian Lan,\authormark{1} Song Zhang,\authormark{2,3} Yuhang Zhu,\authormark{2} Longyu Qing,\authormark{2} Daimin Li\authormark{2,4} Wei Wang\authormark{2,*}, Li Chen\authormark{1,*}, and Wei Wei\authormark{1,*} }

%\address{\authormark{1}College of Optoelectronic Engineering, Chongqing University, 400044, Chongqing, China\\

%\address{\authormark{2}Colledge of Physical Science and Technology, Sichuan University, Chengdu 610064, China\\
	
	%\address{\authormark{3}National Laboratory of Solid State Microstructures, College of Engineering and Applied Sciences and Collaborative Innovation Center of Advanced Microstructures, Nanjing University, 210093, Nanjing, China\\

%	\address{\authormark{4}Institute of Atomic and Molecular Physics, Sichuan University, Chengdu 610065, China\\

%\email{\authormark{*}w.wang@scu.edu.cn} %% email address is required



\begin{abstract}
A high-performance wide-angle refractive index sensor based on a simple one-dimensional (1D) metallic deep nanogroove array with high aspect ratio is experimentally fabricated and demonstrated. The 1D deep groove array is featured by the excitation of magnetic plasmon (MP), referring to an effective coupling of incident electromagnetic waves with a strong magnetic response induced inside the deep grooves. Utilizing the MP resonances that are extremely sensitive to the surrounding dielectric medium, we successfully achieve a refractive index sensitivity (RIS) up to $\sim$1300 nm/RIU, which is higher than that of most experimentally designed plasmonic sensors in infrared region. Importantly, benefiting from angle-independent MP resonances with strong confinement of magnetic field inside the deep grooves and strong electric field localization at the groove openings, we demonstrate wide-angle sensing capability valid in a broadband infrared region with an excellent linear dependence on the change of refractive index. Such a MP-based sensor, together with its simple 1D flat nature and ease of fabrication, has great potential for practical design of high sensitive, cost-effective and compact sensing devices.
\end{abstract}

 % REPLACE WITH CORRECT OCIS CODES FOR YOUR ARTICLE, MINIMUM OF TWO; Avoid using the OCIS codes for “General” or “General science” whenever possible.
%For a complete list of OCIS codes, visit: https://www.osapublishing.org/oe/submit/ocis/

%%%%%%%%%%%%%%%%%%%%%%% References %%%%%%%%%%%%%%%%%%%%%%%%%
\bibliography{ldm_sensor_OE}
\bibliographystyle{osajnl}

%%%%%%%%%%%%%%%%%%%%%%%%%%  body  %%%%%%%%%%%%%%%%%%%%%%%%%%
\section{Introduction}
Surface plasmons, optical excitations with evanescent fields tightly confined at a metal/dielectric interface, are known to be extremely sensitive to the refractive index of the dielectric medium~\cite{Stockman2004,Lal2007,Brongersma2010,Vasa2013}. This unique property is the basis of many fascinating applications such as chemical, biochemical and biomedical sensing~\cite{Kravets2013,Ament2012,Roy2008,Brolo2012,Kabashin2009,Rodrigo2015,Acimovic2014,Im2014,De2011,Anker2008}.

Conventional plasmonic sensors are generally categorized into two types, which are based on surface plasmon polaritons (SPPs) and localized surface plasmons (LSPs), respectively~\cite{Homola1999,Willets2007,Murray2007}. Both SPP and LSP sensors measure the shift of plasmon resonant wavelength to small changes in refractive index adjacent to the metal/dielectric interface. From this point, the sensitivity, which is defined as the resonance shift per refractive index unit (RIU), is generally considered to evaluate the performance of plasmonic sensors~\cite{Mayer2011}. 
LSP-based sensors, supported by noble metal nanocrystals or nanoparticles, are widely used by producing resonant scattering and extinction at specific frequencies, which are featured by extremely strong local field confinement and flexible frequency tunability~\cite{Anker2008,Jackson2004,Li2005,Nie1997}. This enables many applications, particularly in bio-sensing for real-time small molecule detections with rapid and precise tracking of the local refractive index change due to the adsorption of target molecules or bimolecular interactions~\cite{Mayer2011,Anker2008,Brolo2012}. However, LSP sensors are limited to low sensitivity generally on order of $\sim10^{2}\SI{}{\nano\meter}$/RIU, due largely to the small plasmon-active adsorptive interface of nanoparticles. In contrast, SPP-based sensors excited by a prism or metallic gratings (periodic array of grooves, slits, or holes) provide sensitivities that are one or two magnitude higher than that of LSP sensors~\cite{Kabashin2009,Liu2018}. Recent reported plasmonic sensor has achieved a record of sensitivity up to $3\times10^4\SI{}{\nano\meter}$/RIU by integrating a 2D gold diffraction grating with hyperbolic metamaterial~\cite{Sreekanth2016}, which is promising for the applications in clinical diagnosis. Whereas, bulky and sophisticated structure configurations of those plasmonic sensors make them difficult to integrate into cost-effective and portable devices for rapid bioanalytical detections. Therefore, it is highly desired to have plasmonic sensors that rely neither on bulky coupling optics nor on complicated geometrical design.

Analogous to surface plasmons, strong magnetic response, also termed magnetic plasmons (MPs)~\cite{Liu2009,Tang2011,Chen2015,Yang2016,Li2018,Zhu2018}, can be excited in some metamaterials like split-ring resonators (SRRs)~\cite{Enkrich2005} and paired rods~\cite{Zhang2005}. MPs, originated from the coupling of external electromagnetic waves with a strong counteracting magnetic moment, have been widely employed as artificial magnetic atoms for realizing negative-permeability or negative refractive-index metamaterials~\cite{Lezec2007,Xu2013}. The MP resonance, which is sensitive to the surrounding medium, is also attracting for its potentials in sensing applications~\cite{Liu2010,Chen2017,Zhu2018}. Liu et al fabricated a MP-based RI sensor by employing a 2D gold nanodisk array deposited on a metal-dielectric-metal multilayer cavity~\cite{Liu2010}. A sensitivity of $\SI{400}{\nano\meter}$/RIU was experimentally demonstrated. More recently, another MP sensor working at near-infrared frequency was theoretically proposed by using U-shaped metallic split-ring resonators with higher sensitivity over $\SI{1000}{\nano\meter}$/RIU~\cite{Chen2017}. However, MP sensors reported to date are suffering from bulky configurations and complicated geometrical design~\cite{Liu2010,Chen2017}, which dramatically limits their practical realization and applications.

In this letter, we  give an experimental demonstration of a high-quality refractive index sensing by utilizing the MP resonances excited in a simple 1D metallic groove array. A MP-based RI sensor is fabricated with a sensitivity up to $\sim\SI{1300}{\nano\meter}$/RIU resulting from the MP resonances that are extremely sensitive to the surrounding media. Compared with traditional plasmonic RI sensors, this MP-based sensing device is advantageous due to the following three main features: (i) it exhibits stably high sensitivity over a broadband infrared range, which can be readily tuned by appropriately adjusting the geometrical configurations of the groove. A quantitative description of the sensing property has been theoretically analyzed using equivalent nanocircuit model in our previous work~\cite{Zhu2018}, (ii) the 1D nature and the simple geometrical configuration dramatically facilitate the practical fabrication with high compactness and integrability, and importantly (iii) in contrast to SPP-based sensors, the MP-based sensor proposed here can realize wide-angle sensing since the MP resonances are insensitive to the angle of incidence. All these advantages provide an effective way of designing the highly tunable wide-angle refractive index sensors working in desired wavelength range.
\section{Results and discussions} 

\begin{figure}[t]
	\centering
	\includegraphics[width=\linewidth]{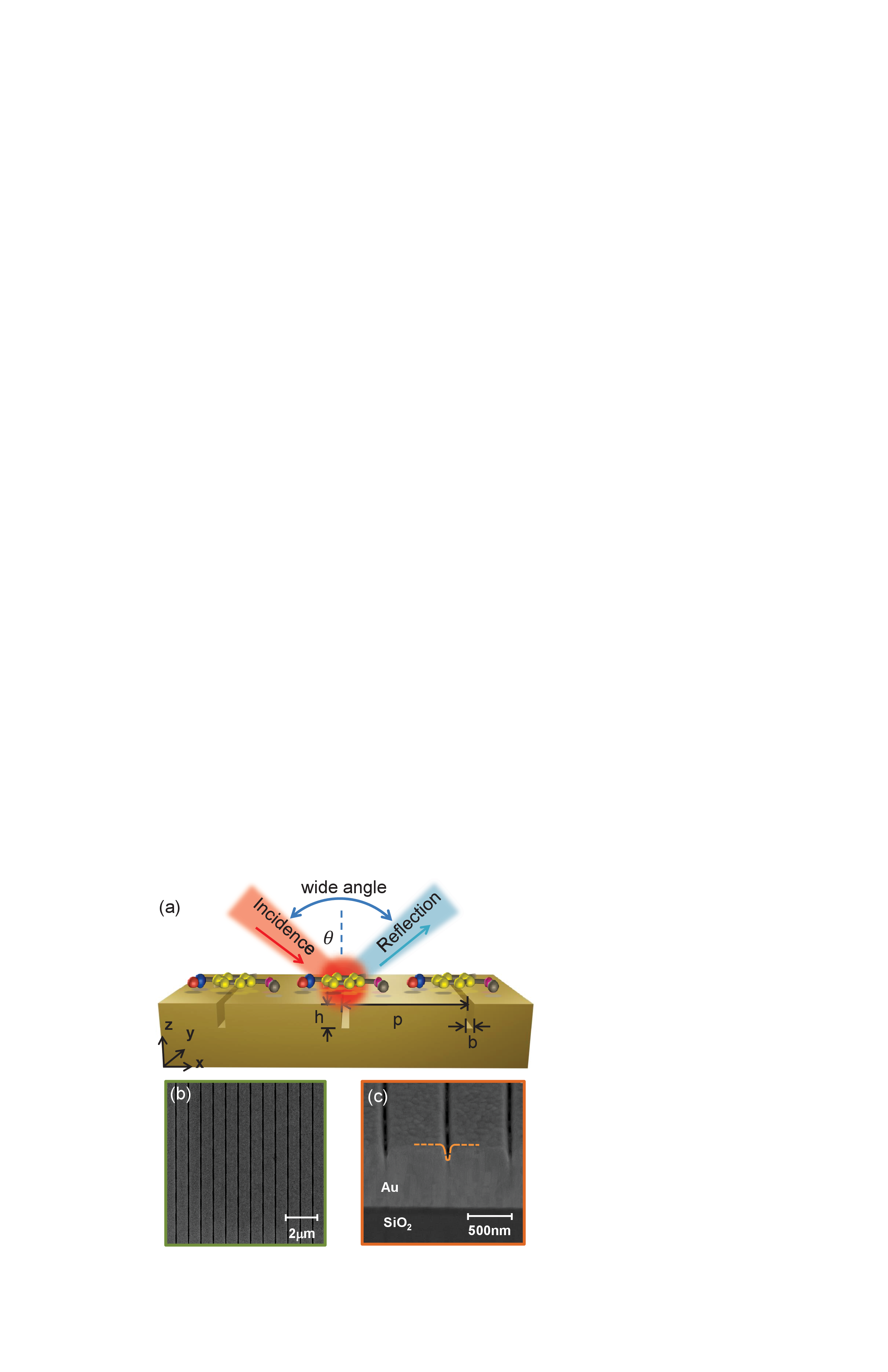}
	\caption{(a) A schematic illustration of gold nanogroove array with high aspect ratio. (b, c) Top- and side-view SEM images of the array with groove period $p=\SI{800}{\nano\meter}$, groove width $b=\SI{70}{\nano\meter}$ and groove depth $h=\SI{200}{\nano\meter}$. The dashed yellow curves mark the groove profile with slightly tapered sidewalls.}
	\label{fig1}
\end{figure}

The sensing device, as depicted in Fig.~\ref{fig1}(a), consists of a periodic nanogroove array featured by deep grooves with high aspect ratio $R=h/b$, i.e., the ratio of groove depth $h$ to groove width $b$. Such deep grooves are the key to excite MPs~\cite{Zhao2014, Zhu2018}. The device is fabricated using focused ion beam (FIB, Helios Nanolab 650, FEI Company) milling of a 500-nm-thick gold film deposited on a silicon dioxide ($\rm{SiO_2}$) substrate. Figure~\ref{fig1}(b) shows the top-view scanning electron microscope (SEM) image of the milled equal-spaced grooves with groove period $p=\SI{800}{\nano\meter}$. A side-view SEM image in Fig.~\ref{fig1}(c) illustrates a clear groove profile, which obviously exhibits a non-rectangular shape with tapered lateral sidewalls, as well as rounding curvatures at both the trench openings and the bottom of the grooves, as marked by the dashed yellow curves. These groove features are the inevitable consequence of FIB (typically with gallium ion source) milling procedure with inherent re-deposition of sputtered material and Gaussian shaped ion beam~\cite{Sondergaard2012,Melli2013}. In order to minimize the deviation from the designed rectangular groove profile with desired depth aspect ratio, an optimized dwell time of $\SI{2}{\micro\second}$ was used and a milling rate was controlled to $\sim\SI{2.3}{\nano\meter}$/pass. A deep groove array with the groove width $b=\SI{70}{\nano\meter}$ and depth $h=\SI{200}{\nano\meter}$ was finally fabricated with high aspect ratio of $R\approx3$.   

Such a deep groove array, under illumination of transverse magnetic(TM) wave with the magnetic component parallel along the groove (y-direction), can induce an oscillating current inside the grooves in the $x{\text{-}}z$ plane. The oscillating current in turn generates a diamagnetic response, which is then coupled to the groove with an excitation of a strong MP  resonance~\cite{Klein2006,Zhao2014,Zhao2014J}.
Similar to SPPs, the MP resonance is extremely sensitive to the surrounding medium~\cite{Zhao2014,Chen2017,Zhu2018}, which makes the nanogroove array a good candidate for RI sensing device. It is worth noting that our simulations (results are not shown here) prove the fact that the slightly tapered groove sidewalls results in small MP resonance shift with respect to the MP excitation in ideal rectangular grooves and small reduction in reflectance amplitude. Whereas, the main feature of MP resonance still holds, which will be illustrated by our simulations in later discussion. 

\begin{figure}[t]
	\centering
	\includegraphics[width=\linewidth]{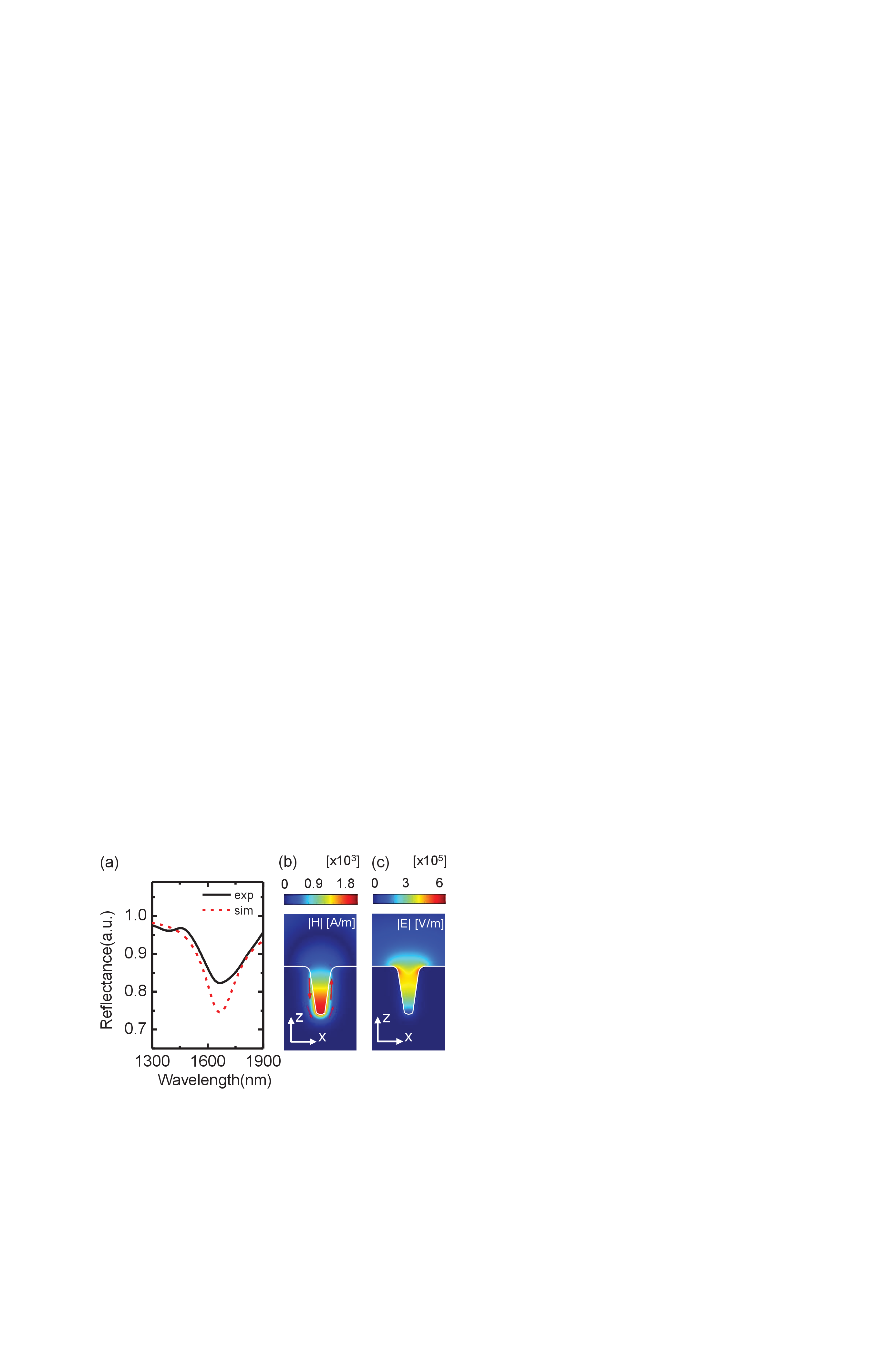}
	\caption{(a) Measured (black) and simulated (dashed red) reflectance spectra at normal incidence for gold array merged in sodium chloride solution with refractive index $n=1.3337$. Simulated amplitude of magnetic field (b) and electric field (c) in color scale at MP resonance $\lambda_{MP}=\SI{1670}{\nano\meter}$. White lines mark the gold-dielectric interface.}
	\label{fig2}
\end{figure}

We first characterized the optical response of the fabricated groove array. Figure~\ref{fig2}(a) gives the reflectance spectrum (solid black) measured on the fabricated nanogroove array covered by sodium chloride (NaCl) solution as surrounding medium with refractive index $n=1.3337$. We can clearly see a prominent dip under illumination of TM wave at normal incidence, corresponding to the excitation of MP resonance $\lambda_{MP}\approx\SI{1670}{\nano\meter}$. We further performed full wave simulations using the commercial finite element(FEM) solver COMSOL multiphysics.  
With optimized geometrical configuration of groove width $b=\SI{93}{\nano\meter}$ and depth $h=\SI{219}{\nano\meter}$, the simulated reflectance (dashed red) was obtained with reasonably good matching to the experimental result, as displayed in Fig.~\ref{fig2}(a). In reality, the milling processing  generates typical non-ideal edges with finite radius of curvature at both the trench and the bottom of the grooves, as clearly marked by the dashed curves in Fig.~\ref{fig1}(c). Therefore, to match the fabricated groove profile, a small non-zero radius of curvature was applied to the corners at both the bottom and trenching opening of the groove, as demonstrated by the white lines in Figs.~\ref{fig2}(b) and (c). We also plotted the distribution of magnetic ($|H|$) and electric field ($|E|$) at MP resonance in Figs.~\ref{fig2}(b) and (c) respectively. Apparently, magnetic field is strongly confined inside the groove, while the electric field exhibits a strong localization at the groove opening, which is a direct evidence for the MP excitation. We will show later that this unique confinement of electromagnetic field plays a key role in realizing high-sensitivity wide-angle RI sensing at infrared frequencies.

\begin{figure}[t]
	\centering
	\includegraphics[width=\linewidth]{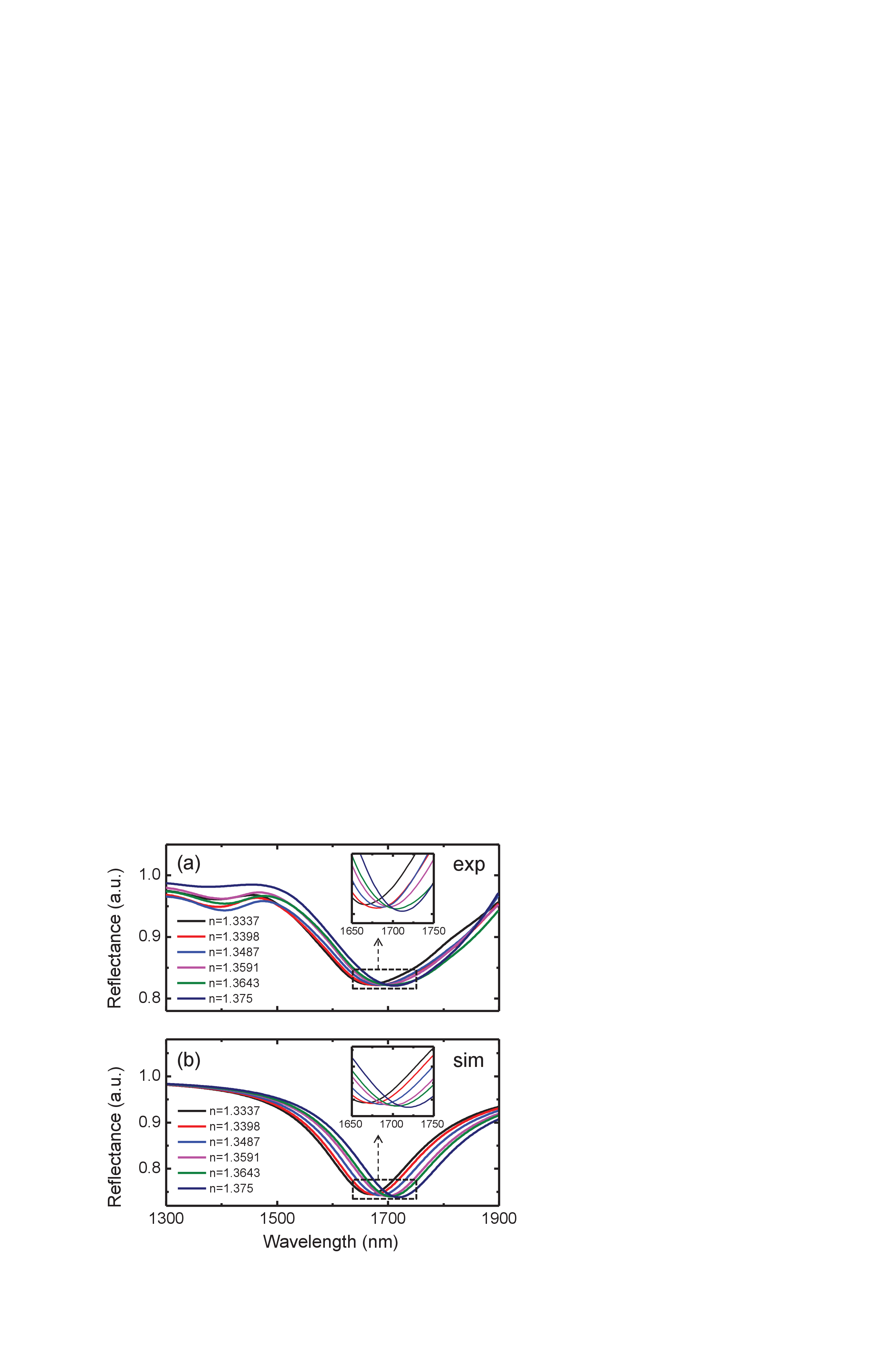}
	\caption{Measured (a) and simulated (b) reflectance spectra of the sensing device with different refractive indices by varying the weight percentage concentrations of sodium chloride solutions. The insets are the zoom-in pictures showing details of the spectra at MP resonances.}
	\label{fig3}
\end{figure}

To demonstrate the potential application of the fabricated nanogroove array as a RI sensor, we utilized a standard sensor evaluation approach to determine the performance of the device. NaCl solutions with various concentrations were injected into the sensor flow microchannel using a syringe to provide minute variations in the bulk refractive index of the groove surrounding medium. The resulting tiny changes in refractive index was recorded as MP resonance shift in the reflectance spectrum. Figure~\ref{fig3}(a) plots the measured reflectance spectra as a function of refractive index of the NaCl solutions. It is apparently shown that the MP resonance is red shifted as the refractive index increases from 1.3337 to 1.375. The inset provides a zoom-in picture for a clearer demonstration of the MP resonance shift, where a considerable shift of $\Delta\lambda\approx\SI{53}{\nano\meter}$ was estimated as a result of a change of refractive index $\Delta n=0.0413$. 
Utilizing the same geometrical parameters as in Fig.~\ref{fig2}(a), we have simulated reflectance spectra for the measured varying refractive index. The simulated results, as shown in Fig.~\ref{fig3}(b), very nicely reproduced the experimental results. 

\begin{figure}[t]
	\centering
	\includegraphics[width=\linewidth]{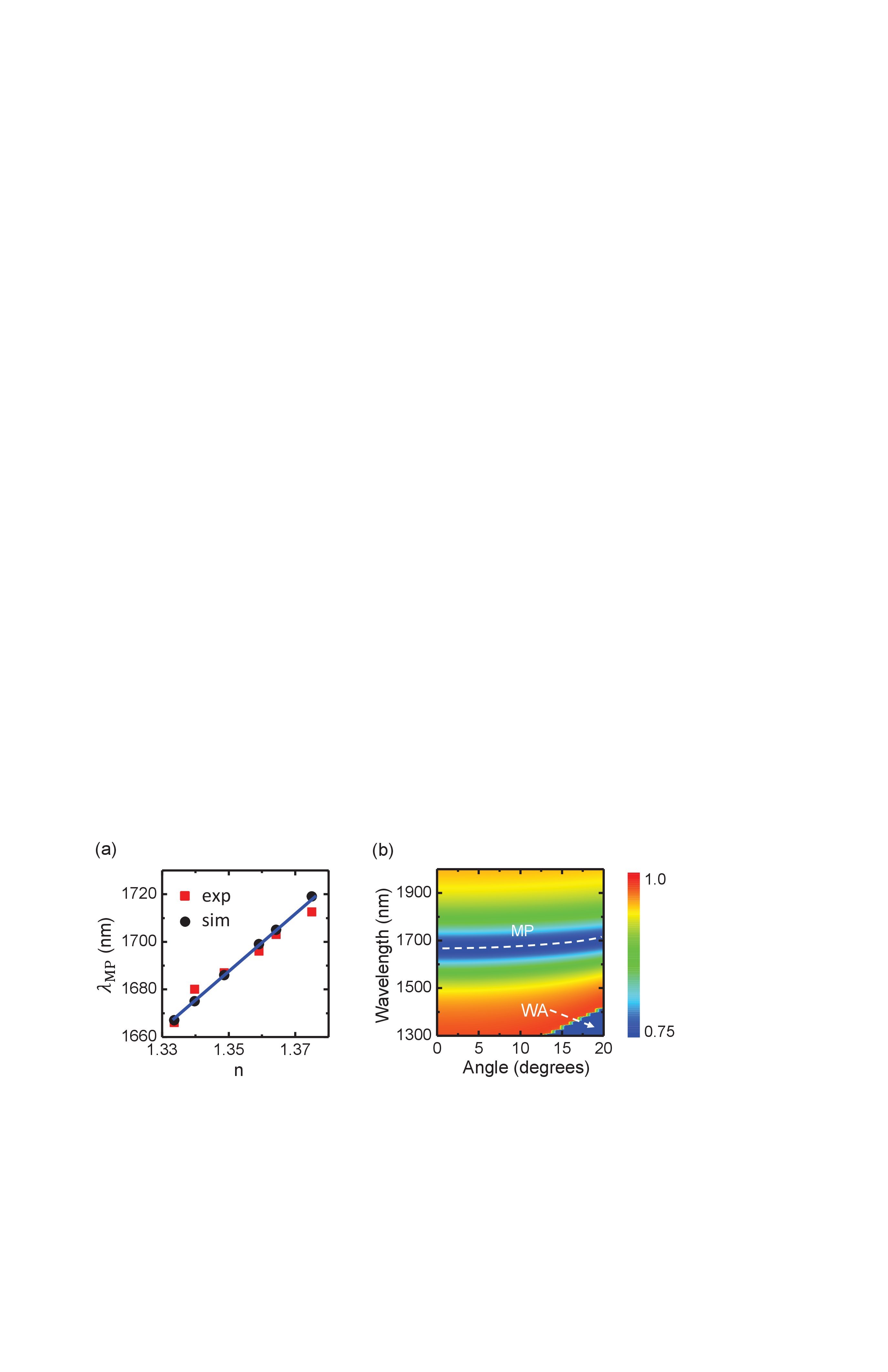}
	\caption{(a) Measured (red symbols) and simulated (black symbols) wavelength positions as a function of refractive indices. The blue line is a linear fit to the experimental data. (b) Simulated reflectance spectra (in color scale) as a function of incidence angle.}
	\label{fig4}
\end{figure}

To quantitatively describe the performance of the sensing device, we have extracted the wavelength position of the MP resonances $\lambda_{MP}$ for different refractive index from the measured reflectance spectra, as shown by the red squares in Fig.~\ref{fig4}(a). Obviously, the measured MP resonance shifts linearly with varying refractive index, which is confirmed by a linear fit (blue line) to the experimental data. This linear dependence has been theoretically predicted in our previous work~\cite{Zhu2018}, where a quantitative description of the changes in MP response excited in the deep grating under the influence of the surrounding medium has been demonstrated by using a modified approach of equivalent inductor-capacitor (LC) nanocircuit.  

In the framework of equivalent LC model, the MP resonance $\lambda_{MP}$ can be expressed in terms of the groove-induced capacitance $C_{\rm{G}}$ and inductance $L_{\rm{G}}$, i.e., $\lambda_{MP}=2\pi c_{0}\sqrt{L_{\rm{Au}}C_{\rm{G}}}$. Here, $c_0$ denotes the speed of light in vacuum. Since $C_{\rm{G}}$ is related to the refractive index of the surrounding medium $n_d$ via $C_{\rm{G}}=Rc_1\epsilon_0 {n}^2_d$ with an invariant $c_1$ being a numerical factor indicating the non-uniform of bound charges, the linear dependence of MP resonance on the refractive index can be readily understood by writing $\lambda_{MP}$ in the following explicit form: 
\begin{equation}
\lambda_{MP}=2\pi c_{0}n_d\sqrt{Rc_{1}\epsilon_{0}L_{\rm{Au}}}.
\label{wlsMP}
\end{equation}
Here, the frequency-dependent gold inductance $L_{\rm{Au}}$ is a slowly varying quantity, which can be regarded as a constant in the interested wavelength range. Equation~\ref{wlsMP} clearly gives the MP wavelengths that are linearly proportional to the refractive index $n_d$ of the filling material. This theoretical predication is further confirmed by our simulated MP resonance extracted from Fig.~\ref{fig3}(b), as given by the black dots in Fig.~\ref{fig4}(a).

In our case, considering the sensitivity, since the MP resonance peak shift is linearly proportional to the change of the refractive index, the first derivative of $\lambda_{MP}$ over $n_d$ is approximately equivalent to a certain amount of MP resonance shift over the corresponding changes in the refractive index:
\begin{equation}
S=\frac{d(\lambda_{MP})}{d(n_d)}\approx\frac{\Delta(\lambda_{MP})}{\Delta (n_d)}.
\end{equation}
Here, we take $\Delta (n_d)$=1.3750-1.3337=0.0413, which corresponds to the peak shift $\Delta(\lambda_{MP})=\SI{53}{\nano\meter}$ from $\SI{1667}{\nano\meter}$ to $\SI{1720}{\nano\meter}$. In this sense, a sensitivity as high as $\SI{1283}{\nano\meter}$/RIU is obtained, which is higher than he observed values in most SPP-based plasmonic sensors reported to date. 

It is important to note that the optical response (the absorption) reported in Ref.~\cite{Liu2010} is much stronger than the current work. However, the corresponding sensitivity value ($\sim\SI{400}{\nano\meter}$/RIU) is surprisingly low. This may be due to the fact the nearly $100\%$ absorption in that multi-layer structure arises from the thin dielectric $\rm{MgF_{2}}$ spacer, in which electromagnetic energy can be efficiently confined. The magnetic field is strongly localized inside the spacer right beneath the gold disks, which is very similar to the case in our previous work~\cite{Zhang2016}. The electric field is also confined under the disks, but closer to the edge of the disk at the gold/$\rm{MgF_{2}}$ interface. Part of the electric field is extended to the air region around the edge of the disk. When evaluating the sensor, air on top of the structure was replaced by glucose solutions with different concentrations. This means that the electric field that really “feel” the RI changes is not sufficiently strong. Whereas, in our case, the electric field is strongly localized at the trench opening and extending into to the groove, as clearly shown in~ Fig.~\ref{fig2}(c). The NaCl solutions not only cover the whole structure surface but also fill inside the grooves. The electric field effectively “feel” the RI changes. Therefore, we assume that the electric field that is involved in sensing is stronger in our case. This may explain our higher sensitivity value even with relatively lower absorptance.

Now we discuss the influence of angle of incidence $\theta$ on the sensing performance of the fabricated nanostructure. One of the most attracting features of MP resonance supported by the deep grooves lie in the fact that it is a strong magnetic resonance induced by the external time-varying magnetic field. Such a diamagnetic response has been widely investigated in the study of metamaterials~\cite{Liu2009,Tang2011}and plays a very important role in realizing negative permeability, or negative refractive index~\cite{Lezec2007,Xu2013}.  
In our case, the incident oscillating magnetic field parallel to the grating grooves (in y-direction) can cause
anti-parallel circulating current around the groove, as displayed by the red arrows in Fig.~\ref{fig2}(b). As the angle of incidence changes, the direction of the magnetic field of the incident light remains unchanged and it can efficiently drive the circulating current for a broad angle range~\cite{Liu2010}. This is in contrast to plasmonic sensors based on propagating SPP, particularly excited by periodic metallic nanostructures, which shifts strongly with different angle of incidence due to its intrinsic angle-dependent dispersion~\cite{Vasa2013,Wei2014,WangJoP_2014,Wei2017,Zhang2018}. This insensitivity of MP resonance to the angle of incidence is beneficial for sustaining the sensing capability for desired frequencies over a broad angle of incidence. 

We have performed simulations for the groove array with embedding material $n_d=1.3337$ by varying angle of incidence from $0^\circ$ to $20^\circ$. Figure~\ref{fig4}(b) gives the simulated reflectance spectra (in color scale) as a function of $\theta$. Two main features are apparent: (i) the MP resonances around $\SI{1670}{\nano\meter}$ with a slight red-shift at higher angle of incidence, as clearly marked by the white dashed line, and (ii) the other broad resonance (dashed arrow) appearing at shorter wavelength, which shifts with varient $\theta$ and is identified as Wood anomaly (WA)~\cite{Hessel1965,Mar2016}. In principle, the MP resonance, as discussed above, is angle independent, while in present groove array, the broad WA resonance shifts to red and starts to interact with the MP resonance as $\theta$ increases up to $15^\circ$. For even larger angle of incidence, the WA and MP resonance will merge together and can not be distinguished (results are not shown). Therefore, the presence of WA resonance dramatically diminishes the sensing performance. 
If we appropriately choose the geometrical parameters of the groove array in such a way that the WA and MP resonance are spectrally far apart, a wide-angle high-sensitivity sensing can be realized in infrared region.

\section{Conclusion} 
In summary, we have experimentally demonstrated a high-quality MP-based RI sensor using a simple 1D array of gold nanogrooves. A sensitivity up to $\sim$1300nm/RIU has been obtained due to the MP resonances that are extremely sensitive to the surrounding media. We have also experimentally shown a stable high-performance sensing ability at infrared frequencies, which results from the linear dependence of the MP resonance shift on the changes in refractive index.  Since the working frequency of the sensors can be precisely predicted by equivalent LC model and readily tuned by simply adjusting the geometrical parameters of the groove array, such a MP-based sensor, together with its simple 1D flat nature and ease of fabrication, is expected to have great potential for practical design of cost-effective sensing devices with high sensitivity and compactness.  

\section{Funding}
This work was supported by the National Natural Science Foundation of China (Grants No.61675139, No.11474207) and the National Key R\&D Program of China (2017YFA0303603).

\end{document}